# Planning horizons and end conditions for sustained yield studies in continuous cover forests

*Jerome K Vanclay*

*Forest Research Centre, Southern Cross University*
*PO Box 157, Lismore NSW 2480, Australia*
*JVanclay@scu.edu.au*

**Abstract**

The contemporary forestry preoccupation with non-declining even-flow during yield simulations detracts from more important questions about the constraints that should bind the end of a simulation. Whilst long simulations help to convey a sense of sustainability, they are inferior to stronger indicators such as the optimal state and binding conditions at the end of a simulation. Rigorous definitions of sustainability that constrain the terminal state should allow flexibility in the planning horizon and relaxation of non-declining even-flow, allowing both greater economic efficiency and better environmental outcomes. Suitable definitions cannot be divorced from forest type and management objectives, but should embrace concepts that ensure the anticipated value of the next harvest, the continuity of growing stock, and in the case of uneven-aged management, the adequacy of regeneration.

**Introduction**

What planning horizon is appropriate in simulation and optimization studies to establish claims of sustainable forestry? This question is vexed enough for even-aged plantations, but becomes even more complex for continuous cover forests managed through natural regeneration (Pommerening and Murphy 2004). Given the attention paid to the broader issue of sustainable forestry since 1992 (e.g., Aplet et al 1993, Maser 1994, Oliver 2003, Higman et al 2004, Espach 2006), and to the technical aspects of simulation modelling and operational research (e.g., Bettinger and Chung 2004, Pretzsch et al 2006, Weintraub and Romero 2006, Bettinger et al 2010, Weiskittel et al 2011), it is surprising that this question about the length of the planning horizon has not been examined more closely, and that there is not more agreement amongst researchers and practitioners. This review examines current norms, and seeks to establish guidelines for further research on planning horizons and aspects affecting yield prediction and planning of forest estates, particularly those practicing continuous cover forestry.

How can one resolve an appropriate length of simulation study to establish that a proposed harvest is sustainable? The answer to this question depends in part on the applicable definition of sustainability and whether a constraint for non-declining even flow is required. A more helpful interpretation of sustainability arises from Hartig (1795) who argued that foresters should utilize forests fully, in a way that future generations will have at least as much benefit as the present generation. This is effectively the same as Bruntland (1987) who expressed the same concept as "meeting the needs of the present without compromising the ability of future generations to meet their own needs". Both these views can be simplified in a forestry context as an objective to maximize current harvests (and services), without impairing future options. Curiously, the contemporary forestry practice to maintain a non-declining even flow seeks neither of these long established goals, hampers the ability to maximize current utility (e.g., to increase harvesting during buoyant markets and to defer harvests during recessions), and does not explicitly seek to avoid impairing the future options (e.g., may not preclude depletion of standing capital). The more simple case of the optimal even-aged rotation has been well studied (e.g., Newman 2002), but the more complex question of the sufficient simulation to demonstrate sustainability warrants further attention.

**Historical precedents**

Evelyn (1664) recorded (chapter 32, paragraph 13) that "*...in Germany and France ... the King's Commissioners divide the woods and forests into eighty partitions, every year felling one of the divisions, so as no wood is felled in less than fourscore years. And when any one partition is to be cut down ... every twenty foot leave a good, fair, sound and fruitful oak standing ... the acorns which take root in a short time furnish all the wood again...*". In this ideal situation, where the site is homogeneous and the climate unchanging, where regeneration is adequate, and the forest is in a steady-state condition, then a one-year simulation is sufficient to prescribe a steady-state harvest. A one-year simulation suffices in this case, because this hypothetical forest is already in a steady-state condition, and because the end condition is precisely defined (as "*eighty partitions, every year felling one of the divisions, so as no wood is felled in less than fourscore years*"). The real world is rarely so convenient, and it is more common to find forests far from steady-state, and to find the identification of steady-state capricious. Thus it is useful to consider a simple theoretical case to shed some light on the way forward.

**Theoretical construct**

Consider a simple case, such as unicellular algae in a jar of water, and assume that its state (e.g., biomass) can be measured with a univariate indicator, $S$. We expect $S$ to follow a characteristic sigmoidal yield curve, and its first derivative, the growth rate, to have a simple maximum such as a quadratic curve (Figure 1).

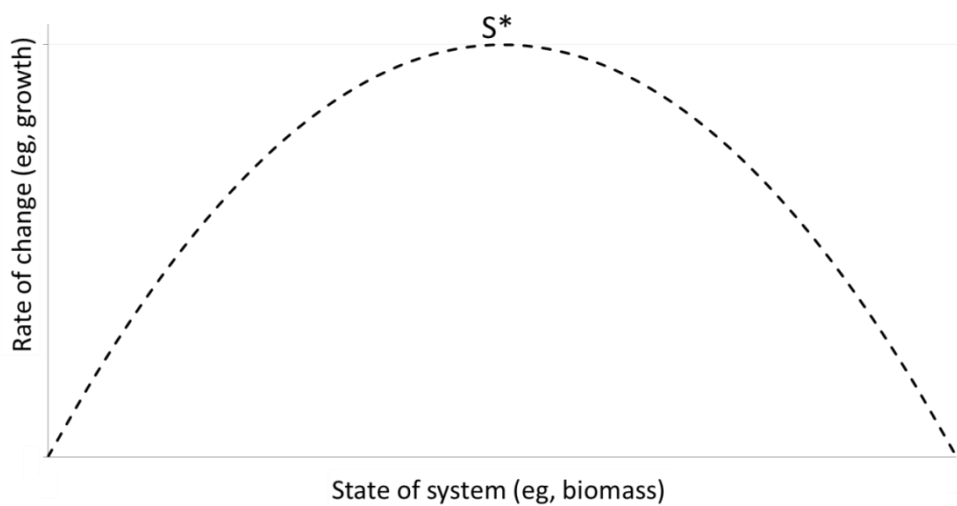

**Figure 1**. Hypothetical response curve typical of many natural resources, indicating how production varies with state of the system, and how an optimal state can be identified.

If the system remains at the optimum ($S^*$) then the net growth $S'$ can be harvested in perpetuity, and a single-cycle simulation may be sufficient to demonstrate sustainability. If the system is overstocked ($S_0 > S^*$), then some harvesting in excess of $S'$ will return the stand to the optimal state, and a simulation of a few cycles may be warranted to demonstrate the return to the optimal state. If the stand is understocked ($S_0 < S^*$), any harvest may need to be deferred or reduced to allow the system to recover to the optimal state, which may require a longer simulation to illustrate convincingly. This theoretical example is overly simple, but helps to indicate that the length of simulation cannot be divorced from the state of the system. This simple example forms the basis for the successful educational game FishBanks (Meadows 1992) which is well respected for teaching sustainability (e.g., Ruiz-Perez et al 2011).

The corresponding analysis for a real world system quickly becomes more complex than the simple theory above. Even the eighty partitions of even-aged oak forest (Evelyn 1664) add much complication: for instance average age is insufficient as an estate-level state indicator ($S$), and mean annual increment (S/n) rather than the current annual increment ($S'$) should be used as the indicator of volume growth (Assman 1970). In addition, because forestry usually involves periodic substantial harvests rather than small annual harvests at any particular site, the optimal post-harvest stand will not coincide exactly with peak production $S'$. And because management objectives for most forests are more complex than the simple goals of Evelyn's example (e.g., Westoby 1987, McKinnell et al 1991, Lawrence and Stewart 2011), the optimal state (S*) should not be viewed purely as a timber goal, but in terms of the productive potential of all the goods and services desired from the forest. Despite this complexity, the enduring principle is the use of simulation studies to assist in maximizing current harvests, without impairing future options. And these simple examples help to illustrate that a short simulation is sufficient for forests close to steady-state, whereas a long simulation is warranted for forests far from steady-state, or for which an optimal state $S^*$ cannot be defined. However, researchers need to be mindful that long simulations are helpful only if they predict and report appropriate indicators: a long simulation displaying only timber yields and omitting other considerations such as biodiversity indicators is unlikely to inform debate about broader aspects of sustainability, particularly if they are unable to offer evidence that the forest remains unimpaired at the end of the simulation.

It seems appropriate to suggest that the appropriate planning horizon is the shortest possible that demonstrates the attainment either of steady-state $S^*$, or of an improved future condition $S_0 < S_n \leq S^*$ (or equivalently, in the case of an overstocked system, $S_0 > S_n > S^*$). It is entirely possible that steady-state may not be reached within a reasonable (say 100-year) simulation, that it is impractical to adequately define a steady state condition, or even that the desired end condition $S_n$ may not be attainable from the current condition $S_0$. In such cases, a suitable compromise may be to demonstrate that the end condition $S_n$ is not inferior to the starting condition (thus $S_0 < S_n \leq S^*$), retains the same production potential, and forecloses no management options.

In theory, it is possible to assert that forest managers should strive for $S_0 < S_n \leq S^*$ when a system is understocked, and $S_0 > S_n > S^*$ when a system is overstocked, the latter with an inequality because $S_n \equiv S^*$ is likely to be unstable given the variable conditions experienced naturally by most forests. In practice, forest management is more complex because of the challenges of defining, measuring and monitoring forest systems. A few industrial plantation monocultures may have narrow commercial goals that can be monitored adequately using simple indicators such as stand basal area, but most forests have more complex objectives that require multi-criteria management goals not easily reduced to a single univariate state variable. Thus although it is useful to conceptualize and seek an optimal forest condition, the reality is that in practice, both the indicators and optima are fuzzy variables that reflect directions for, rather than destinations in forest management.

**Empirical examples of planning horizons**

A survey of the literature reveals a wide range in the length of simulation chosen to investigate the sustainability or consequences of timber harvesting (excluding studies of species succession), and reveals few explanations for the length chosen. Hoogstra and Schanz (2009) suggested that 15 years is the most distant horizon that is realistic for most foresters, and Ferguson (2013) argued that planning horizons beyond 50 years stretch credulity, a contrast to earlier suggestions (e.g., Botkin 1993) that 400-year studies may be needed to infer sustainability. It is common for the planning horizons of yield forecasts to span 60 (e.g., Howard and Valerio 1996, Harper et al 2007) to 100 years (e.g., McKenney 1990, Vanclay 1994, Rohweder et al 2000, Baskent and Keles 2005) or 2-3 harvesting cycles (e.g., Preston and Vanclay 1988, Weintraub et al 1994, Vanclay 1996), while some studies in tropical forests may deal with intervals as long as 400-500 years (e.g., Huth and Ditzer 2001, Sebbenn et al 2008). Few authors document the reasons for selecting a particular planning horizon other than to denote that their choice is consistent with established practice.

Unfortunately, it is even less common to explicitly compare the terminal condition at the end of the simulation with the initial condition of the forest, one of the key indicators of sustainability that is arguably more important than demonstrating non-declining even flow (Vanclay 1996). Whilst optimization studies of silvicultural options often report the desired post-harvesting condition of a single stand, examples of binding constraints on the end point of estate-level simulations are not readily visible amongst refereed publications.

**Distractions and Diversions**

In recent years, there have been several issues that have distracted rather than contributed to the challenge of identifying, characterising and attaining sustainable forest management. Whilst the thrust of this paper is that the key indicator of sustainability is the post-intervention condition (and potential) of the forest, many researchers have been diverted by subordinate issues such as the ratio of successive harvests, and non-declining even flow. Whilst these concepts have influenced monitoring and policy related to sustainable forest management, they have not contributed to the formulation of criteria for post-disturbance conditions (not only post-harvest, but also post-fire and post-storm conditions), aspects that are arguably more important.

In the early 1990s, there was a flurry of activity concerning the relative size of successive harvests, including the concepts of a sustainable original harvest (in which all harvests are the same as the first) and sustainable disturbance harvest (in which subsequent harvests are the same as the second, e.g. Botkin and Talbot 1992, p.62). However, Botkin (1993, p. 212) subsequently warned that "reliance on three harvests can be deceptive as a way to determine whether a harvesting practice is sustainable", and suggested that 400-year studies may be necessary to infer sustainability. Vanclay (1996) demonstrated that inference from simple indices based on estimates of successive harvests was unreliable and could be misleading, and argued that the "sustainability or otherwise of the harvest of timber and other natural resources can only be gauged if the nature of the residual (post-harvest) resource, its entire ecosystem and its ecological and human dynamics are also considered".

Non-declining even flow is a constraint imposed on timber harvests in National Forests in the USA as required by legislation enacted during the 1970s (Behan 1977). Much has written about this legislation and its implications (e.g., Walker 1977, Lenard 1981, Wilkinson 1984, McQuillan 1989, Howard 2001, Mathey et al 2009), but the implication is that it has enshrined the concept of "equal annual yields forever" (Behan 1977) as a central tenet of sustainability. However, a more thoughtful analysis of resource management reveals that non-declining even flow is neither necessary nor sufficient for sustainable forestry, and that it is preferable to focus on the post-intervention stand condition rather than the sequence of harvests (Vanclay 1995, 1996, Howard 2001). Whilst the intent of mandating non-declining even flow is to ensure forest products for future generations, this could be achieved more reliably through closer attention to post-disturbance conditions, and to the optimal forest condition. In addition, the strict application of non-declining even flow hampers economic efficiencies because it ignores fluctuations in timber demand – a weakness easily but rarely overcome by stipulating a decadal even flow rather than an annual even flow.

Another tangent that has received considerable attention is the allowable cut effect, which occurs when a forest manager pledges improved silviculture (e.g., use of better planting stock, fertilizer, thinning) to increase future harvests, thus increasing the non-declining even flow with immediate effect. Although this is conceptually akin to the relationship between interest rates and annuities, the allowable cut effect has stimulated considerable debate and a voluminous literature (e.g., Schweitzer et al 1972, Binkley 1980, Luckert and Haley 1995), most with little consideration of post-harvesting conditions and the long-term future productivity of the forest.

**Research Needs**

Evelyn (1664) advised the use of a number of "*partitions, every year felling one of the divisions, so as no wood is felled in less than fourscore years*", or more generally, *n* years, where *n* represents the optimal rotation age. This remains a sound principle for even-aged estates such as industrial plantations, and considerable research supports the selection of an appropriate harvesting cycle (Valsta 1990, Newman 2002). Established techniques can be used to derive the optimal rotation length, irrespective of whether the focus is narrowly on commercial timber, or more broadly includes environmental services. It is sometimes incorrectly assumed that such partitions should be contiguous and equal in area, but where the estate is heterogeneous it may be desirable or even necessary to form partitions that differ in area so that the expected yields from each partition remain similar.

The concept can be extended to uneven-aged forests with the number of divisions reflecting the interval between harvests (i.e., the cutting cycle), and in turn, the interval for stands to regrow to their pre-harvest condition. Evelyn (1664) suggested "*every twenty foot leave a good, fair, sound and fruitful oak standing*", and thus implied a shelterwood system (Matthews 1991), but plantation management generally involves clearfelling and replanting even-aged stands. Clearfelling makes it trivial to define the residual post-harvest stand, but the specification and implementation of a suitable definition for uneven-aged management is a non-trivial prerequisite for sustainability if it is to form the end condition applied in simulations. Some researchers have focused on de Liocourt's (1898) quotient and the structure of the stand table (numbers in each size class, e.g., Adams and Ek 1974; or with an assumed distribution, e.g., Bare and Opalach 1987, Pukkala et al 2010), but neither of these is particularly satisfactory in mixed stands where some species are non-commercial. The challenge is to offer a workable definition that ensures the anticipated volume of the next harvest, the continuity of growing stock, and the adequacy of regeneration. It may be that a generic stand table with broad species (e.g., early successional, late successional, and non-commercial species) and size classes (e.g., by 20 cm dbh size classes) may be a suitable compromise.

The definition of an optimal state and suitable end-of-simulation constraints becomes more challenging when other environmental objectives are included. In some instances, habitat objectives are prescribed independently of simulations, and may not be sustainable. For instance, native forests harvesting operations in Australia are often required to maintain a specified number of hollow-bearing trees, but few simulations demonstrate the longevity, fate and consequences of these trees. Such conditions are often appended to guidelines without forming an integral part of simulations studies.

**Conclusion**

It appears that the common preoccupation with the planning horizon may detract from the more difficult and more important questions such as end condition and other indicators of sustainability. Whilst long simulations help to convey a sense of sustainability (albeit subject to assumptions embedded within the model), they are an inferior substitute for stronger indicators such as the optimal state and tests of the adequacy of the terminal state at the end of a simulation. As Leary (1985) stated (in a different context) "what began as an interim solution [i.e., long simulations] to a difficult problem [i.e., defining sustainable harvesting] should not now be called the solution to the original problem". It is not easy to define appropriate end conditions for multiple-use forestry, but attempting to define these objectives will offer useful insights into many aspects of forest management and monitoring. Rigorous definitions of sustainability that constrain the terminal state in simulations, should allow flexibility in the planning horizon and relaxation of non-declining even-flow, allowing both greater economic efficiency and better environmental outcomes. The simple expedient of a long simulation with a non-declining even-flow constraint is economically inefficient and insufficient to ensure sustainability. It is preferable to explicitly define a desired terminal condition to simulate a more flexible harvest that ensures that desired outcome.

**References**

Adams, D.M. and Ek, A.R. (1974) Optimizing the management of uneven-aged forest stands. *Canadian Journal of Forest Research* 4(3):274-287.

Aplet, G.H., Johnson, N., Olson, J.T. and Sample, V.A. (1993). *Defining sustainable forestry*. Island Press, Washington DC, 328 pp.

Assman, E. (1970). *The principles of forest yield study*. Pergamon Press, Oxford.

Bare, B. B., and Opalach, D. (1987). Optimizing species composition in uneven-aged forest stands. *Forest Science*, 33(4), 958-970.

Baskent, E.Z. and S. Keles, 2005. Spatial forest planning: A review. *Ecological Modelling* 188:145–173.

Behan, R.W. (1977). Political popularity and conceptual nonsense: The strange case of sustained yield forestry. *Environmental Law* 8:309.

Bettinger, P. and W. Chung, 2004. The key literature of, and trends in, forest-level management planning in North America, 1950–2001. *International Forestry Review* 6(1):40-50.

Bettinger P, K. Boston, J.P. Siry, D.L. Grebner, 2010. *Forest Management and Planning*. Academic Press.

Binkley, C.S., 1980. Economic analysis of the allowable cut effect. *Forest Science* 26(4):633-642.

Botkin, D.B., 1993. *Forest Dynamics: an ecological model*. Oxford University Press, Oxford. 309 pp.

Botkin, D.B. and Talbot, L.M., 1992. Biological diversity and forests. In N.P. Sharma (ed.) *Managing the World's Forests: Looking for balance between conservation and development*. Kendall/Hunt, Dubuque, Iowa, pp. 47-74.

Bruntland, G. (1987). *Our Common Future*: The World Commission on Environment and Development. Oxford.

de Liocourt, F., 1898. De l'amenagement des sapinières. *Bulletin trimestriel, Société forestière de Franche-Comté et Belfort*, juillet 1898, pp. 396-409.

Espach, R. (2006). When is sustainable forestry sustainable? The forest stewardship council in Argentina and Brazil. *Global Environmental Politics*, 6(2), 55-84.

Evelyn, J., 1664. *Sylva, Or, A Discourse of Forest-trees, and the Propagation of Timber in His Majesties Dominions*. The Royal Society of London.

Ferguson, I.S., 2013. Assessing sustainability in certification schemes. *Australian Forestry* 76: 183-193.

Harper, R.J., A.C. Beck, P. Ritson, M.J. Hill, C.D. Mitchell, D.J. Barrett, K.R.J. Smettem, S.S. Mann, 2007. The potential of greenhouse sinks to underwrite improved land management. *Ecological Engineering* 29:329–341.

Hartig, G.L., 1795. *Anweisung zur Taxation der Forste oder zur Bestimmung des Holzertrags der Wälder*. Heyer, Giessen.

Higman S, Mayers J, Bass S, Judd N, Nussbaum R (2004) *The Sustainable Forestry Handbook*: A practical guide for tropical forest managers on implementing new standards, 2nd Ed. Earthscan, London. ISBN 9781844071180.

Hoogstra, M.A. and Schanz, H., 2009. Future orientation and planning in forestry: a comparison of forest managers' planning horizons in Germany and the Netherlands. *European Journal of Forest Research* 128:1–11.

Howard, T. E. (2001). The forester's dilemma: paradoxes in the criteria and indicators for sustainable forestry. *Bois et forêts des tropiques*, 270(4), 75-84.

Howard, A.F. and J. Valerio, 1996. Financial returns from sustainable forest management and selected agricultural land-use options in Costa Rica. *Forest Ecology and Management* 81:35-49.

Huth, A., and Ditzer, T. (2001). Long-term impacts of logging in a tropical rain forest—a simulation study. *Forest Ecology and Management*, 142(1), 33-51.

Lawrence and Stewart (2011) Sustainable forestry decisions: on the interface between technology and participation. *Mathematical and Computational Forestry & Natural-Resource Sciences* 3(1):42–52.

Leary, R. A. (1985). *Interaction theory in forest ecology and management*. Springer.

Lenard, T. M. (1981). How to get less timber and less wilderness at the same time: Wasting our national forests. *AEI Journal on Government and Society*, 5, 29-36.

Luckert, M.K. and Haley, D., 1995. The allowable cut effect as a policy instrument in Canadian forestry. *Canadian Journal of Forest Research* 25:1821-1829.

Maser, C. (1994). *Sustainable forestry: Philosophy, science, and economics*. CRC Press.

Mathey, A. H., Nelson, H., and Gaston, C. (2009). The economics of timber supply: Does it pay to reduce harvest levels? *Forest Policy and Economics*, 11(7), 491-497.

Matthews, J. D. (1991). *Silvicultural systems*. Oxford University Press.

McKenney, D.W., 1990. Multiple-use planning: An application of FORPLAN to an Australian forest. *Australian Forestry* 53(2): 113-123.

McKinnell, F. H., Hopkins, E. R., and Fox, J. E. D. (1991). *Forest management in Australia*. Surrey Beaty, Chipping Norton NSW.

McQuillan, A.G., 1989. The Problem with Economics in Forest Planning: An Overview at Three Levels. *Public Land and Resources Law Review* 10:55-72.

Meadows, D., 1992. Fishbanks Ltd, Institute of Policy and Social Science Research, University of New Hampshire, Durham, NH.

Newman, D. H. (2002). Forestry's golden rule and the development of the optimal forest rotation literature. *Journal of Forest Economics*, 8(1), 5-27.

Oliver, C.D., 2003. Sustainable forestry: What is it? How do we achieve it? *Journal of Forestry*, 101(5), 8-14.

Pommerening, A., and Murphy, S. T. (2004). A review of the history, definitions and methods of continuous cover forestry with special attention to afforestation and restocking. *Forestry*, 77(1), 27-44.

Preston, R.A. and J.K. Vanclay, 1988. Calculation of Timber Yields from North Queensland Rainforests. Queensland Department of Forestry, Technical Paper No 47. 19 p.

Pretzsch, H., Utschig, H., and Sodtke, R., 2006. Applications of Tree Growth Modelling in Decision Support for Sustainable Forest Management. In: H. Hasenauer (ed.) *Sustainable Forest Management*. Springer, Berlin. Pp. 131-149. ISBN 978-3-540-26098-1.

Pukkala, T., Lähde, E., and Laiho, O. (2010). Optimizing the structure and management of uneven-sized stands of Finland. *Forestry*, 83(2), 129-142.

Rohweder M.R., C.W. McKetta and R.A. Riggs, 2000. Economic and Biological Compatibility of Timber and Wildlife Production: An Illustrative Use of Production Possibilities Frontier. *Wildlife Society Bulletin* 28(2):435-447.

Ruiz-Pérez, M., Franco-Múgica, F., González, J.A., Gómez-Baggethun, E., & Alberruche-Rico, M.A., 2011. An institutional analysis of the sustainability of fisheries: Insights from FishBanks simulation game. *Ocean & Coastal Management* 54(8):585-592.

Sebbenn A.M., B. Degen, V.C.R. Azevedo, M.B. Silva, A.E.B. de Lacerda, A.Y. Ciampi, M. Kanashiro, F.S. Carneiro, I. Thompson, M.D. Loveless, 2008. Modelling the long-term impacts of selective logging on genetic diversity and demographic structure of four tropical tree species in the Amazon forest. *Forest Ecology and Management* 254:335–349.

Schweitzer, D.L., Sassaman, R.W., Schallau, C.H., 1972. Allowable Cut Effect: Some Physical and Economic Implications. *Journal of Forestry* 70(7):415-418.

Valsta, L. T. (1990). A comparison of numerical methods for optimizing even aged stand management. *Canadian Journal of Forest Research*, 20(7), 961-969.

Vanclay, J.K., 1994. Sustainable timber harvesting: Simulation studies in the tropical rainforests of north Queensland. *Forest Ecology and Management* 69:299-320.

Vanclay, J.K., 1995. Sustainable silvicultural systems: lessons from Queensland, Australia. In: Ø. Sandbukt (ed.) *Management of Tropical Forests: Towards an Integrated Perspective*. Centre for Development and the Environment, University of Oslo, Occasional Papers from SUM, 1/95, p. 169-181. ISBN 82-90391-25-0.

Vanclay, J.K., 1996. Assessing the sustainability of timber harvests from natural forests: Limitations of indices based on successive harvests. *Journal of Sustainable Forestry* 3(4):47-58.

Walker, J. L. (1977). Economic efficiency and the National Forest Management Act of 1976. *Journal of Forestry*, 75(11), 715-718.

Weiskittel, A.R., D.W. Hann, J.A. Kershaw and J.K. Vanclay, 2011. *Forest Growth and Yield Modeling*. Wiley.

Weintraub, A. and C. Romero, 2006. Operations Research Models and the Management of Agricultural and Forestry Resources: A Review and Comparison. *Interfaces* 36(5): 446-457.

Weintraub, A., Barahona, F., and Epstein, R. (1994). A column generation algorithm for solving general forest planning problems with adjacency constraints. *Forest Science*, 40(1), 142-161.

Westoby, J. (1987). *The purpose of forests*. Blackwell, Oxford.

Wilkinson, C. F. (1984). Forest Service: A Call for a Return to First Principles. *Public Land Law Review*, 5, 1.